\documentclass{article}

 \usepackage[preprint]{neurips_2026}

\usepackage[utf8]{inputenc} 
\usepackage[T1]{fontenc}    
\usepackage{hyperref}       
\usepackage{url}            
\usepackage{booktabs}       
\usepackage{tabularx}
\usepackage{array}
\usepackage{amsfonts}       
\usepackage{nicefrac}       
\usepackage{microtype}      
\usepackage{xcolor}         
\usepackage{comment}
\usepackage{hyperref}
\usepackage{graphicx}
\usepackage{placeins}

\hypersetup{hidelinks}

\title{Neither Layer Alone: Epistemic Integrity Requires Hierarchical Joint Design for Long-Running AI Agents}

\author{%
  Zhihong Shen \\
  Microsoft\\
  Redmond, WA 98052\\
  \texttt{zhihosh@microsoft.com} \\
}

\begin{document}

\maketitle

\begin{abstract}
Long-running AI agents fail not only when inference fails or tools are underspecified, but when independently evolving model and harness layers change the semantics of belief, capability, and goal commitments across their boundary — a failure class this paper terms \emph{interface volatility}. This paper argues that \emph{agent epistemic integrity} (AEI) must be treated as a first-class architectural constraint, achievable only through joint model–harness design organized around an explicit interface contract. 
The central claim is that the model–harness interface contract is the precondition for joint design; its operational form is a four-level hierarchy — goal validity, action-archetype sequencing, tool-instance selection, and invocation-level failure discrimination — that specifies what the boundary must preserve and what structured outputs the model must return for the contract to hold across levels.
This reframes long-running agent design away from flat action loops and toward contract-preserving control over persistent state. Evaluation and training should therefore derive from the contract itself, testing whether belief, tool, and goal commitments hold across session boundaries and independent layer upgrades.
\end{abstract}

\section{Introduction}

An agent can degrade after a model upgrade even when its memory store, tool schemas, and goal records are unchanged. The model may still be capable; the harness may simply have encoded assumptions about how the prior model behaved — whether it trusted memory, deferred to tool output, preserved goals, or allowed external state to override internal priors. When those assumptions no longer hold, stable schemas do not imply stable meaning. This exposes an architectural asymmetry: the model supplies the dominant knowledge prior, while the harness remains reactive. The failure is not ordinary nondeterminism but loss of epistemic meaning under supposedly compatible updates. This paper calls this failure \emph{interface volatility}: behavioral degradation caused by a change in one layer that becomes visible only through integration with the other.

Improving either layer in isolation does not specify the boundary condition between them. Better models may interpret retrieved beliefs, registered tools, and persisted goals more reliably; better harnesses may provide cleaner schemas and stronger validation. But the model still assigns semantic force to external state, and the harness still persists records whose meaning depends on how a particular model reads, prioritizes, or overrides them. Benchmarks may catch regressions and harnesses may block unsafe actions, yet neither guarantees that knowledge, capability, and commitment remain meaningful across independent layer changes. Task training can therefore leave the granular state required for safe resumption unspecified, unversioned, and untested.

This paper argues that long-running agents require \emph{epistemic integrity} as a first-class architectural constraint, achievable only through hierarchical joint design of model and harness, with an explicit model-harness interface contract that precedes and determines the training objective. Stated in the falsifiable form: a long-running agent system whose model-harness interface does not carry epistemic metadata across sessions cannot maintain epistemic integrity in the long-running regime, regardless of model capability or harness sophistication. This claim is scoped to nontrivial long-running settings in which external facts can become stale, tool invocations can have cumulative side effects, goals can expire or change validity, and model or harness versions can evolve independently. In such settings, the issue is observability: histories requiring different safe continuations may collapse into the same model-visible state.

The argument unfolds through three linked claims. First, failures in memory, tool use, and planning should not be treated as separate defects, but as surface forms of a deeper boundary challenge between model and harness: the challenge of \textbf{knowing, doing, and deciding} under distributed state — where \emph{knowing} concerns the representation and priority of memory, context, internal state, and world state; \emph{doing} concerns capability invocation and accumulated effects, and \emph{deciding} concerns goal validity and commitment over time. Second, it proposes \emph{epistemic integrity} as an architectural principle: the authority of external memory, episodic context, tool outputs, goal records, and in-model priors must be governed by explicit, conditional precedence rules rather than implicit prompting conventions. These rules should be specified, learned, versioned, and audited as part of the model–harness contract, with safeguards for stale, inconsistent, or corrupted state. Third, it defines an operational hierarchy in which goal validity, action-archetype sequencing, tool-instance selection, and invocation-level failure discrimination are mediated by distinct interface artifacts and reward signals. The purpose of this hierarchy is to replace flat action-loop thinking with contract-preserving control over persistent, fine-grained state.

This is stronger than observing that the harness matters. Recent work treats the harness as a reliability determinant; AEI gives that premise a stricter architectural form by specifying what the boundary must preserve, what the model must expose for integrity guarantees to remain stable across state transitions, and why a versioned, model-independent contract must anchor long-running evaluation and training. The contribution is not that the harness matters, but that long-running agent design must be organized around the model–harness interface.

The remainder of the paper is organized as follows: Section~\ref{sec:aei} defines AEI and the Knowing, Doing, and Deciding domains; Section~\ref{sec:evaluation} develops evaluation over persisted trajectories; Section~\ref{sec:interface-contracts} specifies the corresponding model--harness contracts; Section~\ref{sec:training} derives contract-first training signals; and Sections~\ref{sec:related-work}--\ref{sec:conclusion} position the framework relative to prior work and outline empirical next steps.

\section{Agent Epistemic Integrity: A Framework}
\label{sec:aei}
\subsection{The Property}
Agent Epistemic Integrity (AEI) is the architectural property that an agent's \emph{active state} remains sufficiently coherent, inspectable, and correctable for the actions it takes over time. The \emph{active state} is structurally essential and operationally central: it includes not just retrieved facts, but also the system's current uncertainty annotations, its understanding of what capabilities have already been invoked and with what effects, as well as its representation of what it is committed to doing next.

AEI is about calibrated operation under incomplete information. A system can be uncertain and maintain AEI only when it behaves proportionally to that uncertainty. Conversely, a system can violate AEI even when individual outputs appear correct, if it acts on stale, partial, or weakly grounded state as if they were fully reliable. Crucially, AEI is a system-level property, not a model-level guarantee. It cannot be ensured by any individual component, and it cannot be evaluated on the basis of individual outputs. Its value is in directing architectural attention toward how the system represents epistemic state and exposes that representation for correction. 

\subsection{Three Invariant Domains}
The problem space of long-running agentic systems decomposes, at the architectural level, into three invariant domains: \textbf{Knowing, Doing, and Deciding}. They are invariant because any agentic system must solve a version of each: it must maintain a picture of state, act through available capabilities, and decide which goals or actions remain appropriate. A solution in one domain does not substitute for a solution in another.

\textbf{Knowing} is memory and context management — what the agent holds as true about the world, itself, and its history. The challenge in the long-running regime is not storage but coherence over time: detecting when standing beliefs have been superseded, resolving conflicts across evidence arriving at different times, and tracing the provenance of decisions back to their epistemic basis.

\textbf{Doing} is capability management — what actions the agent can take, under what conditions, and what the cumulative effect of prior invocations has been. In the long-running regime, capability availability is not binary, effects are not idempotent, and degradation is frequently latent. The challenge is not merely selecting the right tool, but tracking what the system has already done with it.

\textbf{Deciding} is planning and deliberation — how the agent selects and sequences actions given its beliefs and capabilities. In the long-running regime, this extends beyond step-level planning to goal lifecycle: whether the goals that initiated a task remain valid, and whether the agent has a surface for human correction before it acts on goals that have silently expired or updated.

These domains are analytical lenses on a single runtime loop, not separable subsystems. At the model layer, Knowing, Doing, and Deciding are simultaneous aspects of each inference step: the model draws on parametric knowledge in its weights, episodic information in the context window, and harness-supplied external state such as retrieved memories, tool specifications, capability records, and goal-state records. Their separation becomes diagnostic only when attributing failure modes, and architectural only at the harness layer, where memory stores, tool executors, goal-state stores, and orchestration logic can be separately implemented. AEI therefore requires an explicit knowledge-priority contract across all three domains: the interface must specify, version, and audit when harness-supplied runtime state, session context, or parametric priors are authoritative, rather than assuming a fixed precedence order among them.

Table 1 organizes the three domains against two deployment paradigms. The session-based column reflects what current architectures implicitly assume — and encodes the assumption each cell silently makes. The long-running column is where those assumptions fail, surfacing new architectural primitives that session-bounded designs do not require and therefore do not provide. These three domains are not independent — AEI's central claim is that their coupling, not any single domain, is the primary failure surface.

\begin{table}[t]
\centering
\footnotesize
\renewcommand{\arraystretch}{1.25}
\begin{tabularx}{\textwidth}{@{}lXXX@{}}
\toprule
\textbf{Domain} & \textbf{Session-Based} & \textbf{Long-Running} & \textbf{Primitive Required} \\
\midrule
Knowing &
Context fresh by session boundary \textit{(freshness assumed)} &
Detect staleness; revise superseded beliefs &
Belief revision \\
\midrule
Doing &
Invocations locally bounded \textit{(idempotency assumed)} &
Track cumulative state; reason about safe resumption &
Capability state management \\
\midrule
Deciding &
Goal stable for session duration \textit{(goal validity assumed)} &
Track goal validity over time; surface intent drift &
Goal lifecycle management \\
\bottomrule
\end{tabularx}
\caption{The AEI grid. Session-based systems structurally assume freshness, idempotency, and goal validity; long-running operation violates those assumptions and requires belief revision, capability-state management, and goal lifecycle management.}
\label{tab:aei-grid}
\end{table}

\subsection{Prospective Memory as the Unifying Primitive}
All three long-running primitives — belief revision, capability state management, goal lifecycle management — share a common structural requirement: the agent must maintain a live, queryable representation of its own intended future state. This is the function of prospective memory.

The term is borrowed from cognitive psychology, where prospective memory refers to remembering to perform an intended action at a future time \citep{brandimonte1996prospective}. In AEI, a minimal prospective commitment record should expose the intended action or subgoal, status, validity conditions, dependencies, linked beliefs and tool effects, provenance, review or expiry time, uncertainty flags, and revision history. The point is not a canonical schema but the interface obligation: expose what the agent intends to do, why the intention remains valid, what prior state it depends on, and how a human or later model version can revise it. Prospective memory does not replace belief revision or capability-state management; it is the forward-facing integration surface through which revised beliefs and committed side effects constrain what the agent should do next.

Because it exposes forward intentions before consequences occur, prospective memory is the natural surface for human steering. Steerability is therefore not only a user-interface property; it is a memory-architecture property. These commitments are interpreted by a model and persisted by a harness, so AEI treats the interface contract as an integrity surface: the contract specifies what state the harness supplies, what structured commitments the model returns, and which priority rules persist across model upgrades.

\subsection{The Stable What / Volatile How Principle}

AEI specifies the stable what: the invariant obligation to keep beliefs coherent, capability effects resumable, and goals valid across time. The mechanisms that satisfy those obligations -- retrieval strategy, uncertainty scoring, capability logging, or goal-state schema -- are the volatile how. Systems organized around the stable what can survive model and infrastructure evolution; systems organized around today’s mechanisms require redesign when those mechanisms change. This distinction motivates the four-level hierarchy below: the hierarchy fixes the obligations the interface must preserve, while leaving their implementation open to change.

\FloatBarrier
\subsection{Four-Level Hierarchy}
AEI then replaces flat ReAct-style action loops with a four-level hierarchy.

\begin{figure}[t]
    \centering
    \includegraphics[width=1.0\linewidth]{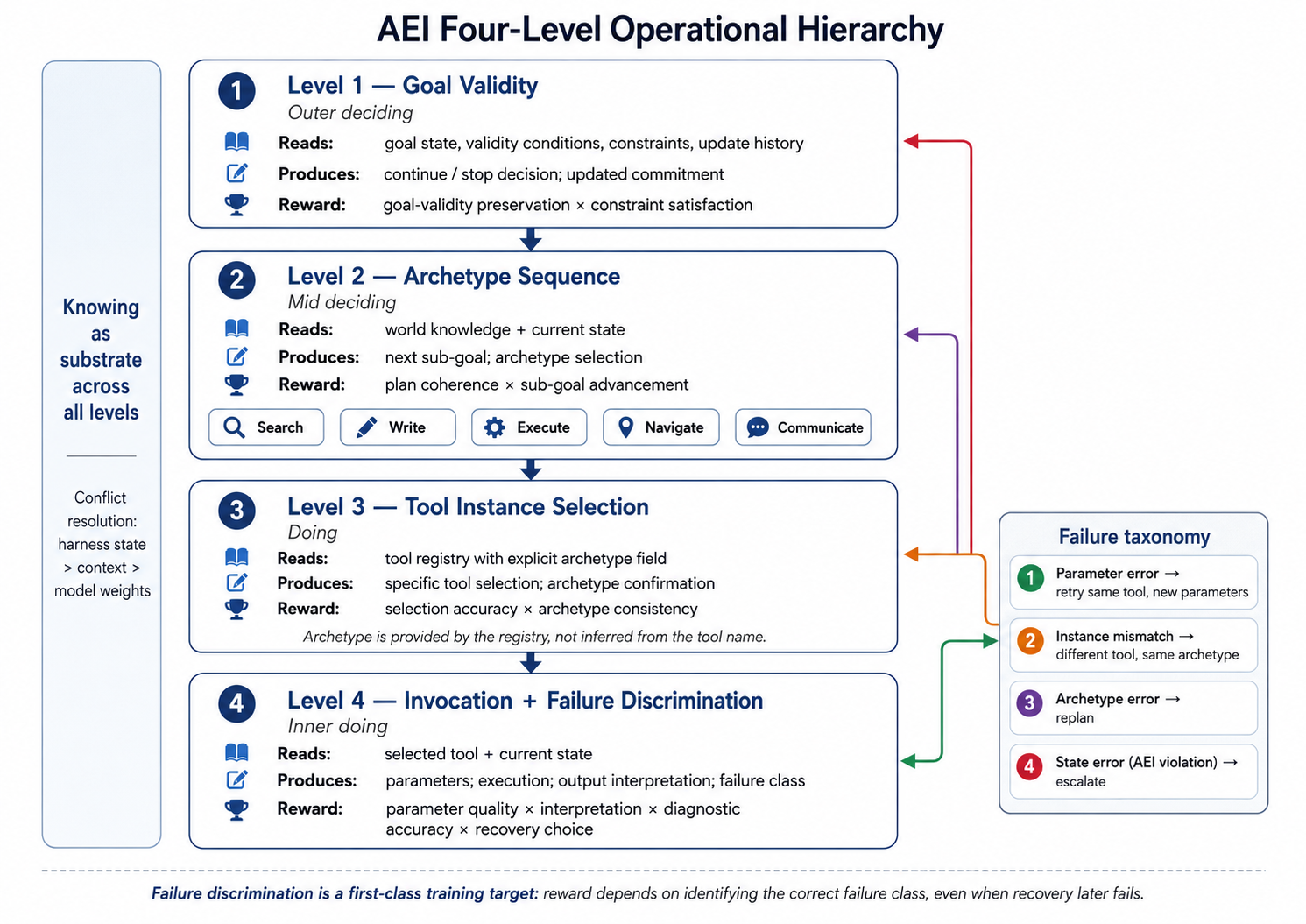}
\caption{Four-level AEI operational hierarchy. The model checks goal validity, selects an archetype, chooses a registered tool instance, and invokes it while classifying failures for retry, reselection, replanning, or escalation.}
    \label{fig:AEI-Four}
\end{figure}

\paragraph{Level 1: goal validity.} The agent asks whether the goal is still valid. It reads the goal state, validity conditions, constraints, and update history. It produces a continuation decision. The relevant reward is not task completion alone. It is preservation of goal validity under constraints. 

\paragraph{Level 2: action-archetype sequencing.} An archetype is an abstract action class that groups tools by the operation they support, before a specific tool instance is chosen. This paper uses search, write, execute, navigate, and communicate as one practical archetype set. The set is not exhaustive, and can be extended as new tool families or agent settings require. At this level, the agent decides the next sub-goal and the action type needed to advance it. It selects an archetype first, then delegates concrete tool selection to Level 3. By separating action type from concrete tool selection, Level 2 makes the model–harness boundary more stable: the model can learn archetype selection independently from registry churn, while later stages train instance selection and invocation with separate reward signals.

\paragraph{Level 3: tool instance selection.} The agent chooses the specific registered tool within the selected archetype, grounded in a registry that explicitly records archetype rather than leaving it to model inference. Its reward is selection accuracy and archetype consistency.

\paragraph{Level 4: invocation and failure discrimination.} The agent synthesizes parameters, executes the tool, interprets output, and classifies failures. A parameter error calls for retrying the same tool with different parameters. An instance mismatch calls for a different tool within the same archetype. An archetype error calls for replanning at Level 2. A state error indicates an AEI violation and calls for escalation to Level 1. Failure discrimination is central to Level 4. Many agents are trained to recover from failure. Fewer are trained to diagnose the kind of failure. AEI treats failure discrimination as its own training target. The reward should measure whether the model identified the right failure class, even if recovery later fails.

This hierarchy is compatible with the Options Framework, which formalizes temporally extended actions through initiation, execution, and termination conditions \citep{sutton1999options}. Here the mapping is architectural rather than fully formal: archetypes initiate from sub-goal needs, execute through tool selection and invocation, and terminate on sub-goal completion or escalation. The contribution is hierarchy at the model–harness boundary: tool use becomes a structured commitment that remains interpretable across state transitions and model versions.

\section{Evaluating Long-Running Epistemic Integrity}
\label{sec:evaluation}

Current trajectory evaluations are useful for bounded task competence: they measure whether an agent completes a task over an observed trace of reasoning, tool calls, planning choices, and recovery behavior. They do not fully test AEI, because AEI failures are cross-session and cross-layer. They emerge when state persists beyond the context window, the harness carries records forward, the model changes while the harness remains fixed, or the tool registry evolves while goals remain active. A benchmark may inspect the final answer and intermediate trace yet still miss the failure if it lacks persistent epistemic state: state that ages, survives resets, crosses model versions, and re-enters later sessions with metadata. AEI evaluation therefore requires reward signals derived from interface obligations across Knowing, Doing, and Deciding, not only local trajectory success.

Cross-session failures can appear in each domain: a memory record may be unchanged while a new model weighs it differently against context; a tool schema may remain valid while the model shifts its inferred archetype or invocation strategy; and a goal record may persist while the model changes how it interprets the goal’s constraints or validity. In all three cases, the harness supplies the same surface state, but the model–harness boundary no longer preserves the same operational commitment.

AEI-aware evaluation should not measure only bounded task success; it should measure whether persisted epistemic state remains stable, interpretable, and safe to resume along five minimum dimensions. The domain labels below indicate the primary stress tested by each dimension, not an exclusive mapping:
\begin{itemize}
    \item \emph{Epistemic drift rate} \textit{(primarily Knowing)}: the fraction of active beliefs whose model-used status diverges from oracle validity after gaps, world updates, or model changes.
    \item \emph{Resumption fidelity} \textit{(primarily Doing)}: whether completed actions, remaining actions, side effects, and retry safety are recognized after interruption.
    \item \emph{Replanning quality} \textit{(Knowing--Doing--Deciding)}: whether revised plans respect persisted constraints, updated facts, and prior tool effects.
    \item \emph{Goal drift detection} \textit{(primarily Deciding)}: precision and recall for expired, contradicted, or invalidated goals.
    \item \emph{Handoff fidelity} \textit{(Knowing--Doing--Deciding)}: metadata preservation across sessions, agents, or model versions.
\end{itemize}

These dimensions are not proposed as a final taxonomy. They are a minimum evaluation surface for making AEI violations observable. The key shift is from evaluating the session to evaluating the persisted trajectory. A persisted trajectory includes prior state, state updates, provenance and validity metadata, tool commitments, goal-validity conditions, and cross-session continuation. Under this view, an agent may complete a local task and still fail AEI if its beliefs are unsourced, its tool choices are not archetype-consistent, its goals are no longer valid, or its failure states cannot be diagnosed and resumed safely.

\section{Model--Harness Interface Contracts} 
\label{sec:interface-contracts}

AEI requires explicit interface contracts across Knowing, Doing, and Deciding. These contracts specify what the harness must provide to the model, what the model must return to the harness, and which parts of the interaction must remain stable across model versions, context resets, and harness changes. 

The need for AEI metadata is an observability constraint, not merely an interpretability preference. Consider two persisted histories that expose the same surface instruction to the model: “email the vendor after confirming approval.” In one history, approval is current; in another, approval was superseded after the prior session. If the interface omits source, timestamp, supersession, and goal-validity metadata, both histories are identical from the model's perspective while requiring different safe continuations. Any model policy receiving only that observation must act identically in both cases and must therefore fail in at least one. The same argument applies to tool side effects and goal commitments. AEI metadata is the mechanism by which semantically different histories remain distinguishable at the model–harness boundary.

A conventional schema specifies fields and types; an AEI contract additionally specifies authority, validity, conflict handling, required model-returned commitments, and invariants that must survive model or harness upgrades.

\begin{table}[htbp]
\centering
\footnotesize
\renewcommand{\arraystretch}{1.25}
\begin{tabularx}{\textwidth}{@{}lXXX@{}}
\toprule
\textbf{Domain} & \textbf{Harness supplies} & \textbf{Model returns} & \textbf{Stable invariant} \\
\midrule
Knowing &
belief, source, timestamp, validity scope, confidence, supersession &
sourced conclusion, uncertainty status, refresh decision &
belief use remains attributable and revisable \\
\midrule
Doing &
tool registry, archetype, capability state, side effects, retry safety &
semantic intent, selected tool, success criterion, failure class &
tool use remains resumable and diagnosable \\
\midrule
Deciding &
goal state, owner, constraints, validity conditions, update log &
continue/revise/escalate, prospective commitment, escalation flag &
commitments auditable/steerable \\
\bottomrule
\end{tabularx}
\caption{Model--harness contract obligations across AEI domains.}
\label{tab:aei-contract}
\end{table}

\paragraph{Knowing contract.}
The Knowing contract governs belief state. The harness must supply beliefs with source, timestamp, validity scope, confidence, and supersession metadata; content alone is insufficient. A fact without provenance cannot be safely revised, and retrieved evidence without source metadata can collapse into ordinary context or the model's parametric prior precisely when it was meant to override that prior. The model must return conclusions with uncertainty and source attribution, distinguishing retrieved evidence from model prior so conflict resolution is visible. This contract addresses Knowing-domain interface volatility by pinning the interpretive frame at the interface.

\paragraph{Doing contract.}
The Doing contract governs tool use. The harness must supply a registry with archetype, capability state, side effects, and retry-safety metadata. The model must return semantic intent, selected tool, success criterion, and failure class with each invocation. The invocation record should state why the tool was selected, what success would look like, and which strategy is being applied. This makes tool use stateful and portable across model or registry changes.

\paragraph{Deciding contract.}
The Deciding contract structures goals and commitments. The harness must represent goals with validity conditions, constraint history, owners, and update logs; a goal statement alone is insufficient because long-running goals can expire, become unsafe, or conflict with later instructions. The model must return prospective commitments in a model-independent, harness-readable format that survives context loss and model upgrades. Under this contract, a goal is a stateful object with explicit validity conditions, not merely text to pursue. This stabilizes goal interpretation without depending on a particular model’s parsing habits.

Across all three domains, the contract defines the stable \emph{what} while models and harnesses may change the volatile \emph{how}; the interface must remain stable enough for continuity. This yields the paper's central claim: the interface contract is not an output of joint design, but its precondition. Without such a contract, rewards cannot target the right behavior, evaluations cannot expose the right failures, harness architects cannot know what state must persist, and model trainers cannot know what outputs must be trained.

\subsection{Tool Registration Semantics: Archetype as Safety Envelope}
The Doing contract’s archetype field is a safety envelope, not a retrieval label: it constrains the effect-semantic properties a tool registered under that archetype may have. Registration must declare two orthogonal properties: \emph{idempotency}, whether repeated identical invocation accumulates state, and \emph{reversibility}, whether side effects can be undone by a compensating operation. The axes are independent: setting a status field is idempotent yet irreversible, while appending a deletable comment is reversible yet non-idempotent. Their combination determines runtime safeguards: non-idempotent irreversible tools require hard confirmation; idempotent irreversible tools require confirmation once and retry only under recorded invocation identity; reversible tools require preserved undo paths. These properties cannot be inferred from names, descriptions, or parameter schemas; the harness must require and enforce them at registration. Read/retrieval archetypes should admit only side-effect-free, idempotent tools; a retrieval tool with cumulative or irreversible effects is misclassified, not merely mislabeled. Write-class archetypes may span risk quadrants, but each tool’s quadrant must be explicit. As registries grow, high intra-archetype variance across the idempotency–reversibility space signals that an archetype has become too coarse or catch-all, degrading taxonomy quality and safeguard assignment — a signal detectable from registration structure alone, without runtime instrumentation.

\section{Contract-First Training: From Task Rewards to Interface Rewards}
\label{sec:training}

AEI changes training design from environment-first to contract-first. A common pipeline builds environments, defines rewards, and trains models, thereby optimizing only what the environment already observes. AEI instead specifies the model–harness contract first, derives rewards from required contract outputs, and then builds environments that generate those signals. This order matters because a model can finish a task without source attribution, use a tool without recording semantic intent, continue a goal without checking validity, or recover from failure without identifying the failure class. Such a model may score well under task objectives while still violating AEI.

The four-level hierarchy implies four reward families: Level 1 rewards goal-validity preservation and constraint satisfaction; Level 2 rewards coherent sub-goal and archetype sequencing; Level 3 rewards tool-selection accuracy and archetype consistency; Level 4 rewards parameter quality, output interpretation, failure classification, and recovery choice. The distinctive signal is failure discrimination: recovery success is not enough unless the model identifies whether a failure is a parameter error, instance mismatch, archetype error, or state error. This requires labeled failure scenarios with diagnostic ground truth.

Under an options view, Level 2 selects the temporally extended archetype while Levels 3–4 supply intra-option execution and failure-discrimination rewards. The exact decomposition is an empirical design choice; the key requirement is that the reward signal preserve the per-level distinctions, especially Level 4 failure-class discrimination, rather than collapsing them into a single terminal or progress reward.

AEI also separates \emph{selection training}—archetype classification and instance retrieval—from \emph{invocation training}—parameter synthesis, output interpretation, and continuation decisions. Treating tool use as a single action hides these distinctions and makes it harder to assign the right reward to the right failure.

Current simulation work provides pieces of this infrastructure: asynchronous environments model dynamic world state, LLM-based simulators generate tool feedback, and code-based world models provide reproducible state machines. AEI adds the cross-session stressor: persistent harness state with aging beliefs, cumulative side effects, goal commitments, and versioned interfaces. Building such environments requires event-driven temporal simulation across session boundaries plus a persistent harness-state layer whose contents are specified by the interface contract.

\section{Related Work}
\label{sec:related-work}

This paper is closest to prior work on epistemic warrant, agent harnesses, tool use, hierarchical agent training, process rewards, uncertainty, and simulated agent environments. Across these areas, existing work improves components of agent reliability; AEI focuses on the contract that must hold across components over time.

Work on trust in artificial epistemic agents asks how AI systems that pursue epistemic goals and shape shared knowledge environments should be evaluated as trustworthy participants in human--AI knowledge ecosystems \citep{marchal2026architecting}. Work on semantic laundering argues that agent architectures can mistake information transfer across trusted tool boundaries for epistemic justification \citep{romanchuk2026semantic}. AEI is complementary to both: it asks what architectural conditions make an agent's belief, capability, and commitment state coherent enough for warrant or trust to be assessed across sessions, model updates, and harness evolution.

Memory-tiering systems in the MemGPT/Letta lineage treat context-window management as a virtual memory problem, using hierarchical memory and paging to make external memory operational for long-running agents \citep{packer2023memgpt}. AEI targets a different problem: not which information is available in context, but which stored beliefs and prospective commitments remain authoritative after world-state changes. Its prospective memory is forward-facing, encoding future commitments and validity conditions for steerability before consequential action.

Recent harness work treats the agent harness as a first-class reliability determinant rather than a thin execution wrapper \citep{meng2026agentharness}. AEI builds on that view by treating the harness as part of the agent's epistemic architecture: the relevant interface is not only an API surface for action execution, but a representational boundary across which beliefs, capabilities, failures, uncertainty, and goals must remain interpretable.

Tool-use and agent-loop work shows that language models can interleave reasoning and acting, call external tools, and route computation through modular components \citep{yao2023react,schick2023toolformer,karpas2022mrkl}. Hierarchical reinforcement learning provides the formal basis for temporal abstraction through options \citep{sutton1999options}, and recent LLM-agent work applies hierarchy to long-horizon planning and credit assignment \citep{peng2026hiper}. AEI uses hierarchy at the model--harness boundary: archetypes define stable tool-use abstractions, while instance selection, invocation, and failure discrimination are delegated to lower-level execution policies.

Process-reward work improves supervision over agent trajectories \citep{xi2025agentprm}, and memory-focused RL improves temporal evidence selection in multi-session dialogue \citep{du2025memoryt1}. These methods are relevant to AEI training, but they primarily supervise local trajectory quality or retrieval correctness. AEI instead requires rewards tied to interface obligations across Knowing, Doing, and Deciding: evidence warrant, capability-state resumption, and goal validity under temporal change.

Simulation work has begun to support dynamic and scalable agent training through asynchronous environments, scalable tool feedback, and executable synthetic worlds \citep{froger2025are,froger2026gaia2,li2025simia,wang2026awm}. AEI emphasizes the cross-session stressors these environments must add: persistent harness state, aging beliefs, cumulative side effects, versioned interfaces, and delayed failures that appear only after model and harness have drifted apart. Control- and uncertainty-oriented approaches such as CAAF and AUQ are complementary \citep{zhang2026caaf,zhang2026auq}: enforcement can constrain invalid actions and uncertainty can guide intervention, but long-running agents still need stable representations of what they know, what they can do, and what they remain committed to over time.

\section{Conclusion and Next Steps}
\label{sec:conclusion}

This paper argues that long-running AI agents require epistemic integrity as a first-class architectural constraint. The central problem is interface stability: when model and harness evolve independently, failures can arise at their boundary even when neither layer is wrong in isolation. AEI reframes this as a joint design requirement across Knowing, Doing, and Deciding, stabilized through explicit model–harness contracts for beliefs, tools, goals, uncertainty, commitments, and failures over time.

The main prescription is that the interface contract must precede training. Rewards should be derived from what the contract requires, and environments should expose whether those requirements remain correct, stable, and resumable. Otherwise, agents can optimize task completion while violating the conditions needed for long-running operation.

The next step is empirical. AEI implies a new class of evaluations and training environments. These should test cross-session persistence, temporal aging, model upgrades, tool-registry changes, and goal-state revision. They should also label failure modes directly, so that models can be trained not only to recover, but to discriminate why recovery is needed.

A minimal testbed would combine three elements: persistent harness state, event-driven temporal change between sessions, and versioned model--harness interfaces. Such a testbed would make interface volatility observable. It would also make epistemic integrity trainable. The deeper point, however, is architectural rather than empirical. The interface contract is not an output of joint design; it is the precondition for joint design. Long-running agent capability does not emerge from independent improvements at either layer alone. It requires contracts that endure the evolution of both.

\bibliographystyle{plain}
\bibliography{references}

\end{document}